# A UV-cured nanofibrous membrane of vinylbenzylated gelatin-poly(ε-caprolactone) dimethacrylate co-network by scalable free surface electrospinning

Mohamed Basel Bazbouz,[1] He Liang,[1,2] Giuseppe Tronci[1,2]

[1]Textile Technology Research Group, School of design, University of Leeds, UK.

[2]Biomaterials and Tissue Engineering Research Group, School of Dentistry, St. James's University Hospital, University of Leeds, UK.

Correspondence: m.b.bazbouz@leeds.ac.uk (M.B.B.), g.tronci@leeds.ac.uk (G.T.)

## Abstract

Electrospun nanofibrous membranes of natural polymers, such as gelatin, are fundamental in the design of regenerative devices. Crosslinking of electrospun fibres from gelatin is required to prevent dissolution in water, to retain the original nanofibre morphology after immersion in water, and to improve the thermal and mechanical properties, although this is still challenging to accomplish in a controlled fashion. In this study, we have investigated the scalable manufacture and structural stability in aqueous environment of a UV-cured nanofibrous membrane fabricated by free surface electrospinning (FSES) of aqueous solutions containing vinylbenzylated gelatin and poly(ε-caprolactone) dimethacrylate (PCL-DMA). Vinylbenzylated gelatin was obtained via chemical functionalisation with photopolymerisable 4-vinylbenzyl chloride (4VBC) groups, so that the gelatin and PCL phase in electrospun fibres were integrated in a covalent UV-cured co-network at the molecular scale, rather than being simply physically mixed. Aqueous solutions of acetic acid (90 vol.%) were employed at room temperature to dissolve gelatin-4VBC (G-4VBC) and PCL-DMA with two molar ratios between 4VBC and DMA functions, whilst viscosity, surface tension and electrical conductivity of resulting electrospinning solutions were characterised. Following successful FSES, electrospun nanofibrous samples were UV-cured using Irgacure I2959 as radical photo-initiator and 1-Heptanol as water-immiscible photo-initiator carrier, resulting in the formation of a water-insoluble, gelatin/PCL covalent co-network. Scanning electron microscopy (SEM), attenuated total reflectance Fourier transform infrared (ATR-FTIR) spectroscopy, differential scanning calorimetry (DSC), tensile test, as well as liquid contact angle and swelling measurements were carried out to explore the surface morphology, chemical composition, thermal and mechanical properties, wettability and water holding capacity of the nanofibrous membranes, respectively. UV-cured nanofibrous membranes did not dissolve in water and showed enhanced thermal and mechanical properties, with respect to as-spun samples, indicating the effectiveness of the photo-crosslinking reaction. In addition, UV-cured gelatin/PCL membranes displayed increased structural stability in water with respect to PCL-free samples and were highly tolerated by G292 osteosarcoma cells. These results therefore support the use of PCL-DMA as hydrophobic, biodegradable crosslinker and provide new insight on the scalable design of water-insoluble, mechanical-competent gelatin membranes for healthcare applications.

# 1. Introduction

In the past decades, electrospun nanofibrous membranes have attracted great attention due to the small size fibres with fine interconnected pores, extremely large surface area to volume ratio and the versatility of polymers, polymer blends and organic-inorganic composite materials that can be smoothly electrospun [1-6]. Electrospun material products have been successfully commercialised as e.g. hernia mesh or vascular access graft, whilst extensive research has been carried towards the design of electrospun therapeutic devices for regenerative medicine [7], tissue engineering, and chronic wound management [8]. While electrospinning is an effective technique for producing nanofibrous membranes, resulting yield of fibre production is typically restricted to 0.1-1.0 g·hour$^{-1}$ for a single spinneret [9], whereby the throughput of the polymer solution in the range of 0.1 to 10 ml·hour$^{-1}$ [10]. Consequently, electrospinning can hardly meet the needs of industrial scale nanofibre manufacture compared with currently available microfibre spinning technologies, enabling the collection of nanofibre nonwoven fabric area of up to 25 cm$^2$ [11, 12]. To overcome this limitation and increase the yield of fibre formation, a great deal of attention has been put towards the development of needle-free electrospinning apparatus, e.g. by Formhals et. al [13] and Jirsak et. al. [14, 15]. Whereby high production of nanofibrous webs could be achieved via the rotation of a roller surface in a polymer solution. Most recently, Elmarco Co. (Elmarco, Liberec, Czech Republic) introduced world's first industrial nanofibre free surface electrospinning set-up i.e. Nanospider® [16]. Here, the polymer solution is electrospun from either wire-based or roller electrodes, so that nanofibre production rate can be conveniently adjusted depending on the electrode width, the linear speed of the wires/roller and the number of spinning heads placed in series [13, 16, 17]. With nanofibrous nonwoven membranes obtained with 50 - 500 nm nanofibre diameter at a production rate of 1.5 g.min$^{-1}$ per meter of roller length [16], this mechanism enables high scalability, low cost, as well as easy operation in comparison with nonwoven membranes electrospun from single spinneret.

Gelatin is a natural biopolymer derived from partial hydrolysis of collagen, mostly composed of randomly-oriented polypeptide chains [18, 19]. Aiming to mimic the nanofibrous architecture of the extracellular matrix (ECM) of biological tissues, gelatin has been successfully electrospun into nanofibrous membranes. In light of its non-toxicity, biodegradability, biocompatibility, formability and low-cost commercial availability [20], gelatin has been excessively used as building block for the design of smart wound dressing and healing materials [21, 22], pharmaceuticals [23], personal care [24] and food industry products [25], as well as drug delivery systems [26-28] and scaffolds for tissue engineering [29]. However, electrospun gelatin typically present uncontrollable water-induced swelling and dissolution, and display weak mechanical strength in the hydrated state, which substantially limit long-term fibre performance [30]. In light of the presence of amine and carboxylic groups along gelatin backbones, various chemical treatments have been proposed to introduce covalent crosslinks lost following collagen extraction and denaturation *ex vivo*, so that micro-/macroscopic structural features and mechanical properties of hydrated gelatin nanofibre membranes could be controlled [31]. Crosslinking strategies have been pursued by carbodiimide chemistry [29, 32-34], bifunctional compounds such as glutaraldehyde (GTA) [33, 35-42] and genipin [30, 43, 44], silanisation [41], dehydrothermal [46] and plasma treatments [47], as well as via ultraviolet (UV) light [30, 48-50]. Although chemical crosslinking is the most widely used method, crosslinking agents are often associated with risks of cytotoxicity and calcification in host polymer scaffolds [51-53], are unable to ensure fibrous retention and minimal membrane dissolution in aqueous media [30, 48, 54], may cause thermal degradation of gelatin [55, 56], or may lead to side reactions, resulting in hardly-controllable process-structure-property relationships. Recently, functionalisation of ECM-derived proteins with photoactive compounds, e.g. 4-vinylbenzyl chloride, has proved to lead to the prompt formation of water-stable, mechanically-competent, UV-cured covalently-crosslinked networks [57], whose preclinical performance has been successfully evaluated in diabetic mice [58]. Together with crosslinking strategies, the use of co-polymers, such as polyvinyl alcohol (PVA) [59, 60], alginate [61], chitosan [62], poly(D, L-lactide-co-glycolide) (PLGA) [63] and poly($\varepsilon$-caprolactone) (PCL) [64-66], has also been pursued as additional means to control gelatin swellability in physiological media [64], with only partial success. Among the different biodegradable polymers investigated, PCL has shown promises as gelatin-stabilising building block, in light of its hydrophobicity, biocompatibility, hydrolytic degradability, whilst it is also FDA-approved for use as degradable suture [45]. At the same time, simple formation of composites made of physically-interacting polymer building blocks is associated with concerns related to the homogeneous material degradability, inevitably leading to material instability *in vitro* and *in vivo* [44, 67]. Here, we explored whether the formation of covalent linkages between the natural and synthetic polymer chains can be achieved to obtain electrospun systems with retained fibrous architecture and enhanced mechanical properties in physiological conditions.

The objective of the study was to investigate whether scalable, geometrically-stable electrospun nanofibrous membranes could be achieved via sequential free surface electrospinning and UV-cured co-polymer network formation in the fibrous state. Vinylbenzylated gelatin was employed as suitable photoactive biomimetic building block and dissolved with PCL-DMA as hydrophobic, hydrolytically-degradable crosslinker. Electrospun membranes were UV-cured aiming to retain their three dimensional and nanofibrous architecture in physiological media via the introduction of covalent crosslinks between polymer chains. The presence of PCL-DMA as additional photoactive phase in the electrospun fibre was expected to mediate the photocrosslinking reaction between distant gelatin chains, thereby leading to enhanced yields of network formation as well as thermal and mechanical properties which respect to electrospun, UV-cured, gelatin controls. The synthesis of the covalent gelatin-PCL co-network was carried out in the fibrous state via UV-initiated radical crosslinking reaction in the presence of (2-hydroxy-4-(2-hydroxyethoxy)-2-methylpropiophenone) (I2959) as water-soluble, UV-compatible photoinitiator [68-71]. I2959 was selected since no toxic response was detected during cell culture with either bovine chondrocytes, human fetal osteoblasts or osteosarcoma cells with up to 0.5 mg·ml$^{-1}$ I2959 [72-74]. To establish defined process-structure relationships, viscosity, surface tension, and electrical conductivity of the electrospinning solutions were measured and linked to fibre characteristics. Scanning electron microscopy (SEM) was pursued to investigate the effect of UV-cured co-network formation on fibres morphology following contact with water, whilst the effect of the solvent employed during UV-curing was also addressed. Differential scanning calorimetry (DSC)-based thermal analysis, attenuated total reflectance Fourier transform infrared spectroscopy (ATR-FTIR) as well as contact angle, tensile and swelling tests were also carried out aiming to identify the structure-property relationships of the electrospun system.

## 2. Materials and methods

*2.1. Materials*

Acetic Acid (AcOH), PCL-DMA ($M_n$: 4000 g·mol$^{-1}$), I2959, 4VBC, 1-heptanol (HpOH), methylene iodide (MI), trimethylamine (TEA), 2,4,6-Trinitrobenzenesulfonic acid solution (5 % w/v, TNBS), Tween-20, N-(3-Dimethylaminopropyl)-N′-ethylcarbodiimide hydrochloride (EDC) and N-hydroxysuccinimide (NHS) were purchased from (Sigma-Aldrich-UK). Phosphate buffered solution (PBS) was purchased from (Lonza-UK).

*2.2. Synthesis of vinylbenzylated gelatin*

Vinylbenzylated gelatin was obtained via reaction of gelatin with 4VBC. Gelatin from porcine skin (type A, high gel strength, Sigma-Aldrich) was dissolved in phosphate buffered solution (PBS, 0.01 M, pH 7.4) via magnetic stirring at 50.0 °C, prior to addition of Tween-20 (1 wt.% of the solution weight). 4VBC was applied to the reaction mixture with a 25 molar ratio with respect to the molar content of gelatin lysines, as determined via (2,4,6)-trinitrobenzenesulfonic acid (TNBS) assay (3·10$^{-4}$ moles of lysine per gram of gelatin) along with an equimolar amount of TEA ([TEA]=[4VBC]). Following 5-hour reaction, the mixture was precipitated in 10-volume excess of pure ethanol; ethanol-precipitated reacted gelatin product was recovered by centrifugation, re-dissolved in PBS and re-precipitated in ethanol, prior to centrifugation and air-drying. The degree of 4VBC-mediated gelatin functionalisation was equal to 40 mol.% of gelatin lysines, as confirmed via 2, 4, 6-trinitrobenzenesulfonic acid (TNBS) [75].

*2.3. Free surface electrospinning and membrane formation*

Vinylbenzylated gelatin (G-4VBC) was dissolved in an aqueous solution of acetic acid (90 vol.% AcOH) at concentrations of 20, 25 and 30% w/v via magnetic stirring at 300 rpm at room temperature for up to five hours. Likewise, PCL-DMA (25% w/v) was dissolved in an aqueous solution of acetic acid (90 vol.% AcOH). G-4VBC (25% w/v) and PCL-DMA (25% w/v) solutions were mixed together to reach an overall polymer concentration of 25% w/v. A gelatin/PCL weight ratio of either (5:1) or (5:2) was selected, corresponding to a molar ratio between 4VBC and methacrylate functions of either (1:1) or (1:2), respectively. All the polymer solutions were left without stirring for five days prior to free surface electrospinning (NS LAB 200 Nanospider Electrospinner, Elmarco, Czech Republic). Either cylindrical or four-wire electrode were used in order to assess and optimise the efficiency of nanofibres formation during free surface electrospinning of both G-4VBC and G-4VBC–PCL-DMA solutions. The applied voltage and electrospinning working distance were 80 kV and 20 cm, respectively. The rotation speed of the electrode and the linear speed of the melt-spun polypropylene (PP) nonwoven fabric collector were (1-5) rpm and 10 cm·min$^{-1}$, respectively. The temperature and relative humidity of the electrospinning environment were 25.0 °C and 30 r.h.%, respectively. The electrospun nanofibre membranes were dried at least 5 days at room temperature to remove any possible remained acetic acid or water before crosslinking.

*2.4. UV-curing of electrospun membranes*

I2959 was dissolved in HpOH at concentrations of either 0.1% or 0.5% (w/v) by magnetic stirring (400 rpm, 25.0 °C) for up to three hours in dark. Previously-obtained electrospun nanofibres were cut into square pieces (60 mm × 60 mm) and fully submerged with I2959-HpOH solution (10 ml) in a 10 cm diameter glass Petri dishes (Sigma-Aldrich, UK). UV curing was initiated with 30-min light irradiation (UVL-16EL, UVP CO. (USA), 365 nm, 0.2 W·cm$^{-2}$) on to both top- and bottom-side of electrospun samples, with a lamp-sample distance of 1 cm. UV-cured samples were washed with HpOH three times to remove any remaining photo-initiator I2959, prior to air-drying for at least 7 days.

*2.5. Viscosity, surface tension and electrical conductivity measurements*

The viscosity of electrospinning solutions (9 mL) was measured on a Brookfield digital viscometer (Brookfield DV-II Viscometer, USA). Readings were taken at 25.0 °C using a spindle and chamber SC4-34/13R, with a shear rate in the range from 60 to 1 rpm (equivalent to 1 – 0.017 sec$^{-1}$). Each sample was equilibrated at the measurement temperature for 10 min before the shear rate was applied.

The surface tension of the electrospinning solutions was analysed by means of the pendant drop method using a CAM 200 contact angle goniometer (KSV Instruments Ltd., Finland). Six parallel measurements were taken using a micro-syringe with an automatic dispenser and a CCD fire wire camera (512x480) with telecentric zoom optics combined with LED based background lighting allowing capturing images at frame intervals from 10 ms to 1000s. The surface tension $\gamma$ was calculated by using equation 1:

$$\gamma = \frac{g \cdot \rho \cdot d_e^2}{S_f} \quad [\text{Eq.1}]$$

Where $g$ is the acceleration of gravity, $\rho$ is the difference in density between air and the polymer solution, $d_e$ is the equatorial diameter of the droplet and $S_f$ is the shape factor, as referenced in [76]. Each measurement was replicated three times per sample and the mean values and the standard deviation were calculated and plotted.

The electrical conductivity of the electrospinning solutions was measured by Novocontrol broadband dielectric impedance spectrometer (Montabaur, Germany), with integrated alpha-A high performance frequency analyser in the frequency range of 0.1-10 MHz. Samples were tightened by using vacuum sealed parallel electrodes (Ø: 18 mm) as a specimen holder cell with a spring for sealing out the air. Data were analysed by using Win Fit software. All measurements were carried out at a temperature of 25.0 °C and the spectrometer was set to a bias of 1 volt. Each sample was measured between frequency sweeps of 1.0 Hz to 1.8 MHz.

## 2.6. Scanning electron microscopy (SEM)

SEM was employed to examine the 2D surface morphology and web structure of the electrospun nanofibres made of either VB-G or VB-G–PCL-DMA before and after crosslinking (SEM; Hitachi S-2600N, Japan) with a secondary electron detector. All air dried samples were placed onto aluminium stubs (Ø: 12.7 mm) with the aid of carbon adhesive tapes and sputter gold-coated (Emitech K550X, UK) under high vacuum. An acceleration voltage of 3 - 4 kV was used with a typical working distance of 6.5-15.0 mm. Images were captured at different magnifications in the range of 200x to 30000x. The average fibre diameters in each sample were estimated by evaluating a minimum of 50 fibres at different regions and the mean values and the standard deviation were calculated and presented.

## 2.7. Differential scanning calorimeter (DSC) analysis

DSC (DSC Q100, TA Instrument, New Castle, USA) was conducted on raw material of G-4VBC and PCL-DMA as well as respective electrospun and UV-cured nanofibrous membranes. A heating rate of 10°C·min$^{-1}$ (Nitrogen flow rate of 30 cm$^3$·min$^{-1}$) and a temperature range of 10-250 °C was applied, whilst 2-5 mg dry sample weight was used in each measurement.

## 2.8. Attenuated total reflectance Fourier transform infrared spectroscope (ATR-FTIR) analysis

ATR-FTIR spectra were acquired on dry electrospun samples of G-4VBC and G-4VBC–PCL-DMA before and after UV-curing via a Perkin Elmer Spectrum BX spotlight spectrophotometer with diamond ATR attachment system. Scans were taken in the range of 4000-600 cm$^{-1}$ and 64 repetitions were averaged for each spectrum sample. The resolution was 4 cm$^{-1}$ and the scanning interval was 2 cm$^{-1}$.

## 2.9. Mechanical properties characterization

Mechanical properties of electrospun nanofibrous membranes were tested in dry and wet conditions by using an Instron single column testing machine 5544 (Instron, Norwood MA, USA) equipped with a 5 N load cell, pneumatic side action grips and *Merlin* materials testing software. All the nanofibrous membranes were cut into rectangles with a size of 25 mm (length) × 10 mm (width). The thicknesses of the samples were measured by using a ProGage digital micrometer thickness tester (Thwing-Albert Instrument Company, west berlin, USA) having a precision of 0.1 micrometer. Each thickness measurement was replicated three times per sample and the mean values were calculated.

For the measurement of the nanofibrous membranes in the wet condition, specimens were immersed in phosphate-buffered saline (PBS, pH = 7.4) at room temperature for 24 h. Tensile tests were carried out (n=3) at a temperature of 25.0 °C and relative humidity of 65% with a crosshead speed of 10 mm·min$^{-1}$ and a gauge length of 20 mm for both dry and wet conditions. The results of the experiments were computed in a load (cN) versus extension (mm) plot. Stress-strain curves were generated by dividing the load by the cross-sectional area of the nanofibre mat and the extension by the gauge length with their conversion factors, respectively, so that the mean values and the standard deviation of the Young's modulus, strain at break and tensile strength were calculated and plotted.

*2.10. Liquid Contact Angle (LCA) measurement*

The hydrophilicity/hydrophobicity of electrospun G-4VBC–PCL-DMA membranes were examined via time dependent contact angle measurements. LCAs were measured at the air/ web surface interface at room temperature of 25.0 °C, and 65% relative humidity, using a pendant drop method through the KSV Cam 200 optical contact angle drop size analyser (KSV instruments Ltd., Helsinki, Finland). Approximately, 10 µL of methylene iodide (MI) was dropped onto the surface of each flat nanofibrous membrane via a gas tight micro syringe with a 27G needle. LCA values were monitored by taking sample images at a rate of one image per each second.

*2.11. Water holding capacity (WHC) measurements*

The water uptake capacity of nanofibrous membrane samples was measured by 24-hour incubation of dry samples (of known weight ($w_d$) and area (3 cm$^2$)) in distilled water at room temperature. Subsequent to sample equilibration, each sample was collected, paper-blotted and corresponding hydrated weight ($w_s$) measured immediately by a semi micro balance (Sartorius CP 225D, Germany). The swelling ratio (*SR*) was calculated as follows:

$$SR = \left(\frac{w_s - w_d}{w_d}\right) \times 100 \qquad [Eq.2]$$

Each measurement was replicated three times per sample and the mean values were calculated and recorded.

*2.12. Cell culture*

G292 osteosarcoma cells were cultured in Dulbecco's modified Eagle's medium (DMEM), supplemented with 10% fetal bovine serum (FBS), 1% glutamine, and 2.5 mg mL-1 penicillin-streptomycin, in a humidified incubator at 37 ˚C and 5% CO2. Cells were passaged every 3 days with 0.25% trypsin/0.02% EDTA. The UV-cured samples were incubated in a 70 vol.% ethanol solution under UV light. Retrieved samples (n=6) were washed in PBS three times, prior to cell seeding. G292 cells (8×10$^3$ cells·ml$^{-1}$) were seeded on top of the membrane samples (following UV disinfection) and incubated at 37 ˚C for 4 days. After incubation, UV-cured samples (n=6) were washed with PBS (x 3) and transferred to a new 24 well plate before adding the dying agent of Calcein AM and Ethidium homodimer 1; the plate was then incubated for 20 minutes away from light. Finally, live and dead stained samples were placed on to a glass slide for fluorescence microscopy imaging (Leica DMI6000 B). Other than live /dead staining, cellular activity was assessed using Alamar Blue assay (ThermoFisher Scientific, UK) according to manufacturer's guidance. Alamar Blue data were presented as percent ratio of the fluorescence intensity recorded in the sample (or control) treated with cells with respect to the fluorescence intensity recorded in the same sample (or control) without cells. Statistical analysis was carried out using OriginPro 8.5.1. Significance of difference was determined by one-way ANOVA and post-hoc Bonferroni test. A p value of less than 0.001 was considered to be significant. Data are presented as mean ± SD.

Together with the UV-cured samples, a crosslinked gelatin control was prepared via state-of-the-art carbodiimide-mediated crosslinking chemistry, as reported previously [77]. Briefly, a 10 wt.% gelatin solution was prepared in distilled water at 50 $^{\circ}$C. An equimolar amount of EDC and NHS was added to the gelatin solution with a 0.8 molar ratio with respect to the molar content of gelatin lysines (~3 ×$10^{-4}$ Lys·$g^{-1}$). Following 1 min stirring at 50 $^{\circ}$C, 1 gram of crosslinking mixture was casted onto each well of a 24 well-plate and incubated at 50 $^{\circ}$C for three hours. Further to extensive washing, resulting films were dehydrated with an increasing series of ethanol solutions in distilled water (0, 20, 40, 80, 100 vol.%) and cultured with cells as previously-described.

## 3. Results and discussion

The manufacture of a UV-cured membrane consisting of a covalent co-network of G-4VBC and PCL-DMA enabling retainable fibrous architecture in aqueous environment is presented. Results will be discussed with regards to FSES process, fibre characteristics, co-network formation, membrane properties and stability in aqueous environment. Sample nomenclature used in this work is: 'GX' and 'GYPZ' identify the electrospinning solutions of either G-4VBC (G) or blends of G-4VBC and PCL-DMA (P), respectively; 'X' indicates the concentration of G-4VBC in the electrospinning gelatin solution, whilst 'Y' and 'Z' refer to the weight ratio (5:1 or 5:2) of G-4VBC (5) and PCL-DMA (1 or 2) in the gelatin/PCL electrospinning solution. Electrospun membranes are coded as either 'F-GX' or 'F-GYPZ', where 'F' indicates the fibrous sample, whilst the other letters have the same meaning as described above. UV-cured membranes are indicated as 'F-$GX^W$' or 'F-$GYPZ^W$', where 'W' identifies the concentration of I2959 photoinitiator used during UV-curing, whereby low (0.1% w/v I2959) and high (0.5% w/v) photoinitiator concentrations were coded as either 'L' or 'H', respectively. PBS-hydrated electrospun membranes are coded as either 'PBS-F-$GX^W$' or 'PBS-F-$GYPZ^W$'.

*3.1. FSES and UV-curing of nanofibrous membranes*

Figure 1 shows the overall research strategy pursued for the manufacture of UV-cured nanofibrous membranes via FSES. Gelatin was covalently-functionalised with 4VBC photopolymerisable adducts, as key chemical functions to mediate the synthesis of a UV-cured gelatin/PCL co-network. PCL-DMA was employed as hydrophobic, degradable crosslinker in the electrospinning mixture aiming to induce photocrosslinking of distant gelatin chains during UV fibre irradiation, so that fibre swelling in aqueous environment could be controlled.

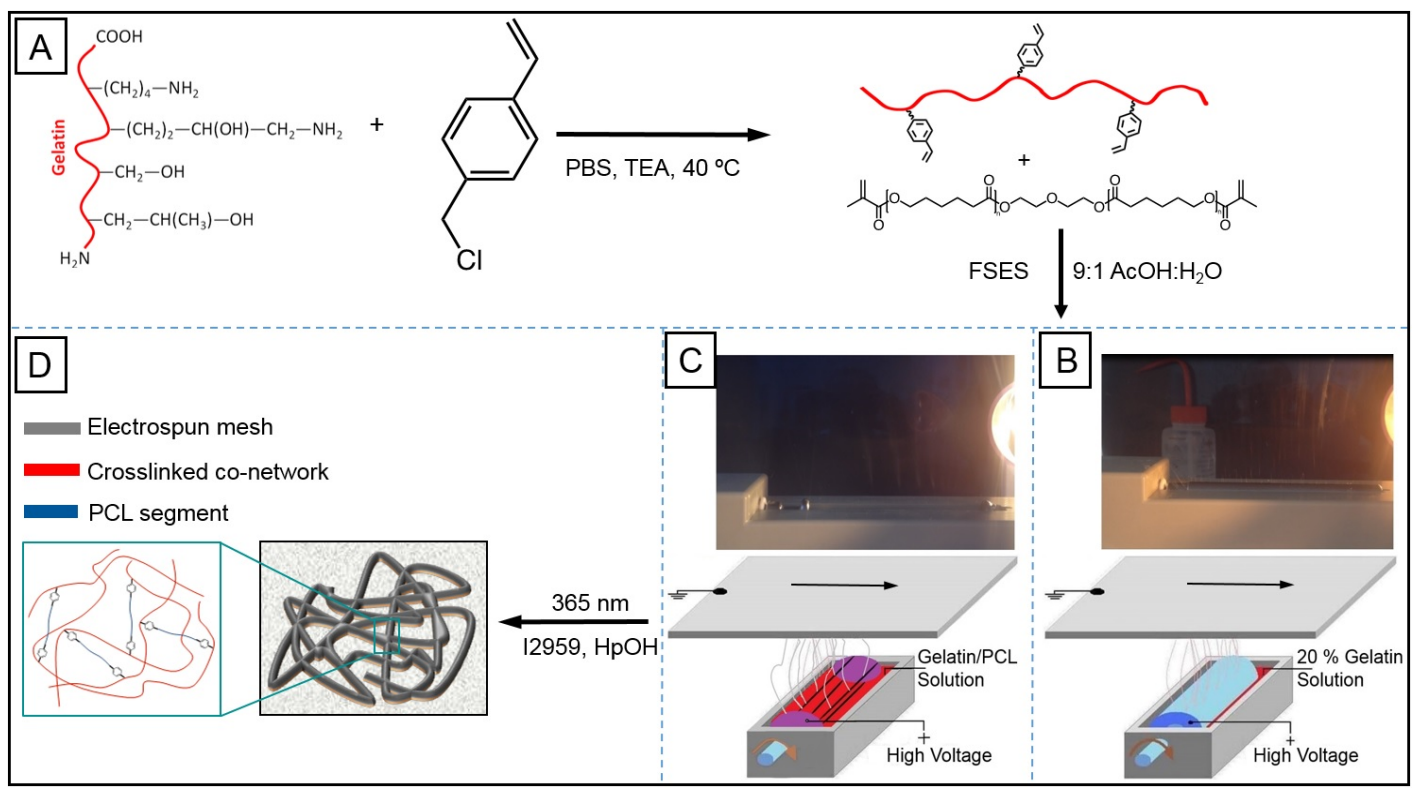

**Fig. 1** FSES-based scalable manufacture of the UV-cured electrospun membrane with enhanced structural stability in aqueous environment. Gelatin is chemically-functionalised with 4VBC (A) and electrospun in the presence of PCL-DMA, resulting in fibres bearing photopolymerisable groups at the molecular scale. FSES set-up features a polymer solution bath (connected to high voltage power supply) and a rotating collection plate (connected to the ground). Fibre-forming electrospinning jets can be obtained with either cylindrical (B) or four-wire electrode (C) following electrode contact with the polymer solution bath. UV irradiation of the electrospun membrane (D) leads to the synthesis of a gelatin/PCL covalent co-network, whereby the presence of PCL-DMA is key to achieve photocrosslinking between distant gelatin chains.

FSES-induced fibre formation was successfully demonstrated with both cylindrical and four-wire electrodes (Fig. 1 B, C), which are partly immersed in the polymer solution bath, facing a collector plate. The collector plate is connected to the ground and carries a meltspun PP nonwoven fabric rotating perpendicularly to the fibre-forming electrospinning jet. As the electrode rotates through the polymer solution bath, a thin film of charged polymer solution is loaded on to either the cylinder or wire surface. Consequent to electrode rotation and application of voltage, the formed film undergoes a Plateau–Rayleigh instability, initially resulting in the formation of droplets [17], and subsequently, in polymer solution jets. Whilst the jet diameter decreases following solvent evaporation, it must be noted that the electrostatic charges on the jet expand the jet in the radial direction and stretch it in the axial direction. Furthermore, the radial forces from the charges become large enough to overcome the cohesive forces of the jet [2]. Based on this concept, as the diameter decreases the radial forces increase to split the jet into two or more charged jets which are approximately equal in diameters and charges per unit length [78]. Each smooth segment that is straight or slightly curved suddenly develops an array of whips. After a short sequence of an unstable whipping back and forth, each jet follows a bending, winding, spiralling and looping path in three dimensions and produces a large number of super fine fibres 'nanofibres' moving towards the collector [10].

Despite above-mentioned mechanism was similar in both set-ups, cylindrical and four-wire electrodes proved to provide increased yield of fibre formation and stability with low and high viscosity electrospinning solutions, respectively, i.e. solution G20 on the one hand, and solutions G30, G5P1 and G5P2, on the other hand. To ensure retention of fibrous structure and photocrosslinking of gelatin chains in the fibre state, electrospun fibres were incubated in a I2959-containing solution of HpOH. HpOH was selected as a suitable non-solvent for both gelatin (to avoid dissolution of gelatin fibres) and water (HpOH solubility: $8 \cdot 10^{-6}$ mol·g$^{-1}$, 1 atm, 25.0 °C [79]), whilst allowing for the complete dissolution of I2959 photoinitiator. The selection of a non-solvent for both gelatin and water was crucial since gelatin fibres were electrospun from aqueous solutions and proved to contain small amount of moisture even after drying due to the hygroscopic nature of gelatin. In the case of fibre incubation with a common non-solvent for gelatin, e.g. ethanol, water diffusion gradients proved to be established between the fibre and surrounding environment, due to the miscibility of ethanol with water, so that fibre instability and merging were observed following contact with ethanol (See Figure S1, Supporting Information). By using HpOH on the other hand, above-mentioned fibre instability mechanisms were minimised in light of the immiscibility of HpOH with water. Consequently, fibre-forming gelatin chains could be fixed in a crosslinked state following UV curing, so that fibrous structure could be retained in aqueous environment.

*3.2. Viscosity, surface tension and electrical conductivity of G-4VBC and G-4VBC–PCL-DMA solutions*

FSES can only be accomplished when specific process parameters are conveniently controlled, such as evaporation rate of the solvent, as well as viscosity, surface tension and electrical conductivity of the electrospinning polymer solution.

Appropriate evaporation rate of the solvent is crucial aiming to get stable fibre formation. Solvents with high evaporation rate (e.g. Hexafluoro-2-propanol (HFIP)) will obstruct and clog the jet formation from the diffused droplets present on either the four-wire or cylindrical electrode. On the other hand, solvents with low evaporation rate will lead to the formation of 'wet' nanofibres on the collection fabric, ultimately resulting in a film of merged fibres rather than a fibrous web. In this study, a solution of acetic acid in distilled water (9:1) (v: v) was successfully selected as electrospinning solvent, resulting in the formation of bead-free gelatin nanofibres, as in the case of traditional needle-based electrospinning set-up [80, 81]. The acidic pH of the electrospinning solvent was beneficial to achieve complete dissolution of both G-4VBC and PCL-DMA at room temperature, without requiring to the application of heat during FSES.

The shear viscosities of G-4VBC solutions at three concentrations (i.e. 20%, 25% and 30% w/v) and G-4VBC–PCL-DMA solutions at two 4VBC/MA molar ratios (i.e. 1:1 and 1:2) were characterised at varied shear rates under constant temperature. As expected, the shear viscosity of the gelatin solutions increased noticeably with the increase in G-4VBC concentration over the full range of shear rate investigated. Gelatin/PCL solutions (prepared at constant polymer concentration) displayed increased viscosity when the 4VBC/MA molar ratio was decreased, consistently with the increased content of PCL in respective solution (Fig. 2 A). Considering the electrospinnability of the system, as the viscosity of the dissolved polymer is increased, the entanglement of the polymer chains will also be increased, until reaching a polymer concentration window allowing for the elongation of the jet to occur and breakup of the ejected jet to be prevented. Further increase in the solution viscosity will, on the other hand, induce suppressed capillary instability, resulting in the formation of beads in the nanofibrous web morphology. All the polymer solutions investigated showed shear thinning behaviour, i.e., the solution viscosity found to be decreased at increased shear rates. That is an indicator that all of them are non-Newtonian liquids [82].

Together with the solution viscosity, the surface tension of the electrospinning solutions was also quantified (Fig. 2 B). Values were observed to significantly decrease as the gelatin concentration decreased, i.e. from nearly 35 mN·m$^{-1}$ (sample G30) to about 23 mN·m$^{-1}$ (sample G20). For the G-4VBC–PCL-DMA samples at concentration of 25% w/v, the surface tension (~ 27 mN·m$^{-1}$) was on the other hand minimally affected by the decrease in 4VBC/MA molar ratio (1:1 → 1:2) and insignificantly different from the one measured in PCL-free gelatin solutions of comparable polymer concentration (sample G25). It is generally considered that the surface tension plays an important role in determining the range of polymer solution concentrations from which continuous uniform nanofibres can be obtained in electrospinning [81], where the viscoelastic forces completely dominate the surface tension. In other words, a direct correlation has been observed between uniform fibre diameter and polymer solution viscosity and surface tension [83]. In comparison, when the polymer solution concentration is significantly reduced, the viscoelastic forces are dramatically reduced and thereby the surface tension plays a strong role in the morphology of the resulting fibres. Hence, when the surface tension forces are dominant, they attempt to reduce surface area per unit mass, and thus beaded fibres are consequently produced [84]. Reported surface tension values of the pure solvents used are as follows, the surface tension of water at 27°C is 71.73 mN/m [85] and the surface tension of acetic acid at 20 °C is 27.61 mN/m [85]. It is clearly seen that the acetic acid concentration in water strongly affected the surface tension of solution, and the surface tension can be reduced by increasing the acetic acid concentration in water [86].

The electrical conductivity as a function of frequency is presented in the (Fig. 2 C) for samples of dissolved gelatin at various concentrations and dissolved gelatin/PCL at two blend ratios. The value of the electrical conductivity at 1 Hz has been considered as a direct current (DC) conductivity for comparative purpose which suits the case of the DC high voltage power supply used in our work. However, figure 2 (C) shows all electrical conductivity curves are increasing as the frequency increased. Therefore, differences between all dissolved gelatin solutions were assessed at the lowest frequency (1.0 Hz, nearly zero). For the dissolved gelatin at various concentration systems the electrical conductivity increases as the gelatin concentration decreased, i.e. from 345 µS·cm$^{-1}$ in G30 to 390 µS·cm$^{-1}$ in G20. Similar trends of electrical conductivity were also observed in gelatin/PCL solutions with decreased 4VBC/MA molar ratios, i.e. from 302 µS·cm$^{-1}$ (sample G5P1) to 258 µS·cm$^{-1}$ in sample G5P2. These results can be explained considering the variations in polymer and acetic acid concentration in the electrospinning solution. The electrical conductivity of a polymer solution is directly related to the electrostatic charge density carried by the electrospinning jet. At constant voltage and solution-to-collector distance, solutions with increased electrical conductivity typically lead to highly-elongated jets and the formation of electrospun fibres with smaller diameter [87].

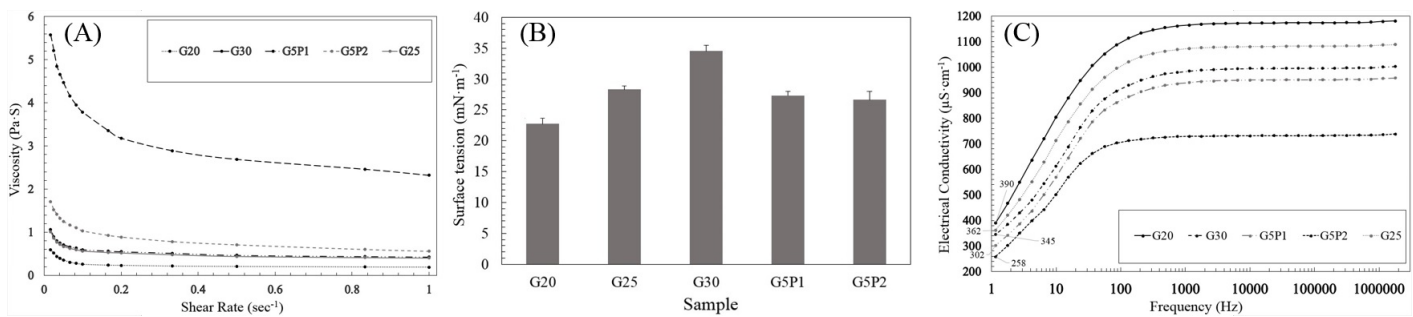

**Fig. 2** Viscosity (A), surface tension (B) and electrical conductivity (C) measurements of G-4VBC solutions at three concentrations (i.e. 20 %, 25 % and 30 % w/v) and G-4VBC–PCL-DMA solutions at two 4VBC/MA molar ratios (i.e. 1:1 and 1:2).

## 3.3. SEM morphology of electrospun and UV-cured membranes

FSES parameters were investigated aiming to electrospin either functionalised gelatin or respective gelatin/PCL solutions into nanofibres with the desired morphology. Typically, electrospun nanofibres with small diameter (several micrometers down to less than 100 nm) and thus high surface area (approximately 3 to 80 $m^2 \cdot g^{-1}$) are beneficial for the manufacture of regenerative devices, aiming to promote cell homing [88-90]; such membranes can typically be electrospun from solutions with decreased polymer concentration, whilst specific electrospinning parameters need to be identified in order to avoid the formation of non-uniform, beaded fibres. SEM images shown in figures 3, 4 and 5 indicate that uniform fibre diameter distribution (140±45 nm to 1.5±0.6 μm) could be achieved from both gelatin and gelatin/PCL solutions by selecting polymer concentrations in the range of 20-30 % w/v, whilst applying an electrospinning voltage of 80 kV, an electrode-to-collector distance of 20 cm and an electrode rotation speed of 1- 5 rpm. Depending on the polymer solution viscosity, the electrode rotation speed was decreased when electrospinning solutions with increased viscosity in order to get stable electrospinning process with continuous nanofibres production. These parameters were kept constant aiming to at the fabrication of uniform gelatin and gelatin/PCL nanofibres, and subsequent UV-induced crosslinking reaction.

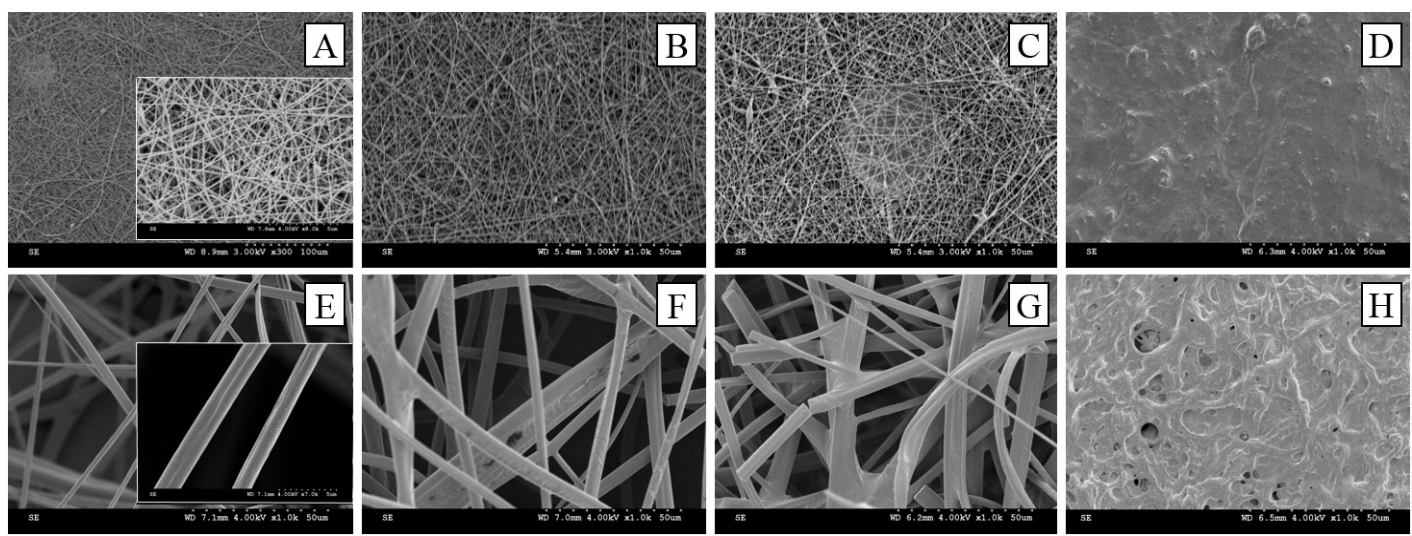

**Fig. 3** SEM images of electrospun and UV-cured gelatin membranes prior to (A-C, E-G) and following incubation (D, H) with water. (A): F-G20; (B) F-G20$^L$; (C, D): F-G20$^H$; (E) F-G30; (F) F-G30$^L$; (G, H): F-G30$^H$.

The morphology of the electrospun and UV-cured gelatin fibres is shown in figure 3. The average diameters of electrospun nanofibres were approximately 140 ± 45 nm and 1.5 ± 0.6 µm, when solutions with polymer concentration of either 20 or 30 % w/v were applied, respectively (Fig. 3(A, E)). Incubation of gelatin fibres in HpOH and subsequent UV irradiation in the presence of I2959 did not induce detectable alteration of fibre morphology, as shown in (Fig. 3(B, C, F, G)), whereby the diameter of individual fibres and the pore size remained mostly unchanged. These observations support the use of HpOH as non-solvent for both gelatin and fibre-incorporated, so that no fibre dehydration and merging could be caused during fibre incubation, in contrast to the case where water-miscible solvents, e.g. ethanol, were applied. Consequently, the structural stability of UV-cured gelatin nanofibre mats was evaluated via 24-hour incubation in water at room temperature, followed by air drying. Respective samples were not dissolved in water, in contrast to electrospun webs, yet respective fibre morphology was significantly changed, resulting in a film of merged nanofibres (Fig. 3(D, H)). Although the synthesis of a UV-cured gelatin network was supported by the macroscopic stability of resulting membrane, the yield of photocrosslinking was suggested to be too low to promote retained fibrous architecture at the microscopic levels.

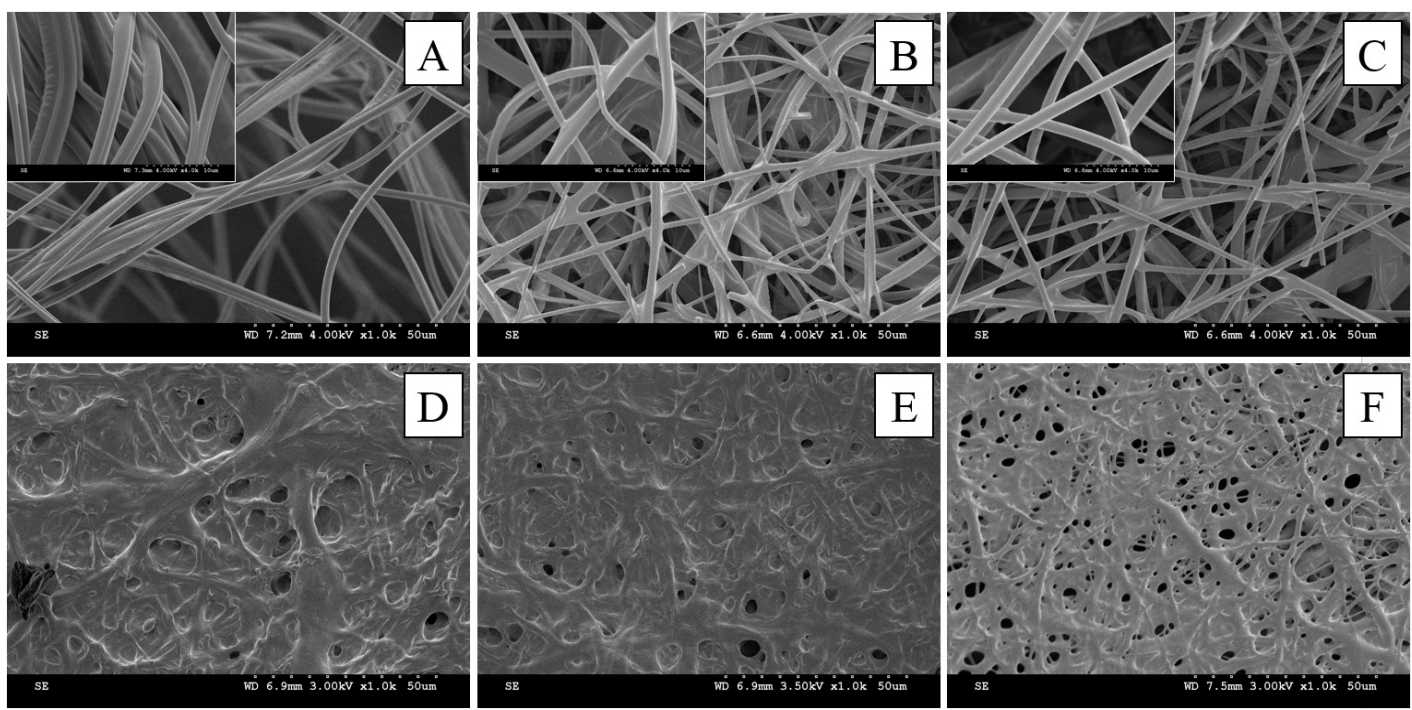

**Fig. 4** SEM images of electrospun and UV-cured samples G-4VBC–PCL-DMA prior to (A-C) and following incubation with water (D-F). (A, D): F-G5P1; (B, E) F-G5P1$^L$; (C, F) F-G5P1$^H$.

Consequently, PCL-DMA was introduced in the electrospinning solution as hydrophobic, degradable crosslinker, aiming to enhance the crosslinking density of resulting gelatin/PCL co-network and to minimise photocrosslinking-hindering steric effects during UV curing [86]. Figures 4 and 5 show the morphology of the electrospun fibrous samples F-G5P1 and F-G5P2 with a 4VBC molar ration of 1:1 and 1:2, respectively. The employment of PCL-DMA in the electrospinning solution did not significantly impact on the fibre diameter, with respect to PCL-free samples electrospun from solutions with comparable polymer concentration, in line with previously-discussed properties of respective electrospinning solutions. Average fibre diameters of 1.2 ± 0.3 µm and 1.4 ± 0.7 µm were measured in electrospun and

UV-cured samples, respectively (Fig. 4 (A, B, C) and Fig. 5 (A, B, C)). Interestingly, the synthesis of the PCL-gelatin covalent co-network in UV-cured electrospun samples proved to result in a lower averaged fibre diameter with respect to the one previously reported in electrospun PCL-gelatin blends [44]. Most importantly, the original fibrous morphology was partly retained in samples F-G5P1$^{L-H}$ following incubation in water (Fig. 4 (E, F)), whilst nearly-retained fibres could be observed in water-incubated samples F-G5P2$^{L-H}$ (Fig. 5 (E, F)). As confirmed via SEM, the increase of both the PCL content in the electrospinning solution and the photo-initiator concentration in the UV-curing HpOH bath was key to achieve membranes with retainable porous and fibrous morphology in physiological environments.

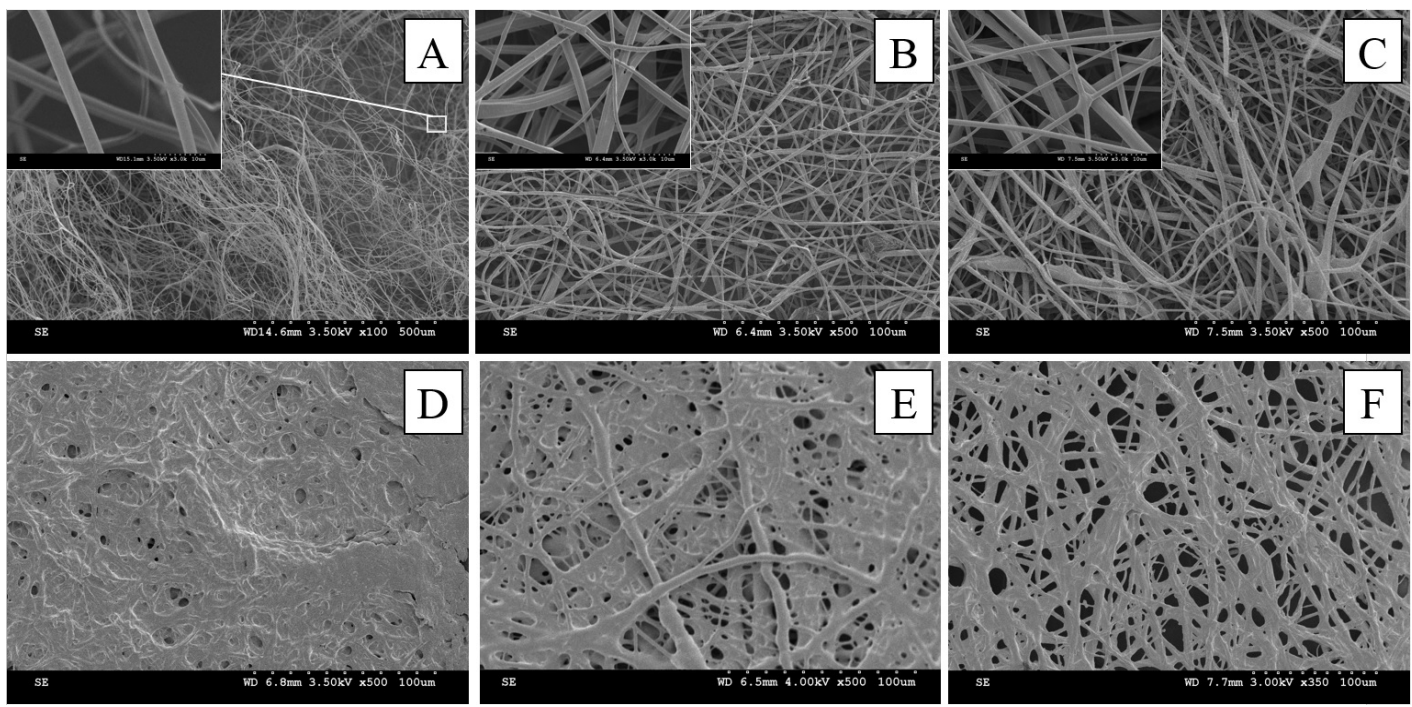

**Fig. 5** SEM images of electrospun and UV-cured G-4VBC–PCL-DMA prior to (A-C) and following incubation with water (D-F). (A, D): F-G5P2; (B, E): F-G5P2$^L$; (C, F): F-G5P2$^H$.

Low magnification scale (x350) scanning electron microscopy (SEM) on sample F-G5P2$^H$ (Fig. 5(F)) therefore demonstrated that UV-cured electrospun samples successfully displayed nearly-retained fibres and increased structural stability following incubation in aqueous medium. Confirmation of wet-state fibre retention at such low magnification scale differentiates this SEM investigation with respect to previously-reported ones [44, 65], whereby the fibrous scaffold morphology following sample hydration was only recorded at high magnification scales in the range of x2000-20000.

*3.4. Characterisation of thermal properties*

Following investigation of fibrous architecture and membrane stability in the hydrated state, the attention moved to the investigation of material properties. Figure 6 shows the differential scanning calorimetry (DSC) thermograms of raw, electrospun and UV-cured samples. Endothermic peaks were recorded at 130 °C and 45 °C for samples G-4VBC and PCL-DMA, corresponding to the denaturation of gelatin ($T_d$) and melting temperature ($T_m$) of PCL, respectively. The value of $T_d$ is in agreement with the value reported in the literature for pristine gelatin [31], suggesting that the functionalisation with 4VBC does not induce any detectable variation in gelatin thermal properties. The value of $T_m$ recorded for PCL-DMA is lower than the typical value displayed by PCL ($T_m$: 60 °C), which is somewhat expected considering the lower molecular weight of PCL-DMA. Electrospun sample F-G30 displayed a $T_d$ at around 97 °C, which is lower than the one recorded in the raw material. The most likely explanation for this finding is that electrospinning of gelatin induces unwinding of any partially-refolded collagen-like triple helices, consequent to the application of high voltage to the gelatin solution, so that decreased thermal stability is observed [87, 91]. Other than PCL-free samples, the endothermic peak of gelatin was found at lower values in thermograms of samples F-G5P1 and F-G5P2 with respect to the case of sample F-G30, whilst PCL-related endothermic peak of $T_m$ could not be detected, likely related to low content of PCL in the electrospun material. It has been indicated that during the preparation of the polymer solution and the electrospinning process, denaturation of collagen-like triple helices can take place in gelatin samples, resulting in decreased thermal stability. These observations are likely to explain why above-mentioned endothermic peaks recorded in electrospun samples of G-4VBC and G-4VBC–PCL-DMA are shifted towards lower temperatures. It has been completely agreed in the literature that the crosslinking treatment will enhance the thermal stability of the electrospun gelatin fibres [31, 35, 48, 87-90]. This is also observed for the UV-cured samples F-G5P2$^L$ and F-G5P2$^H$, whereby values of $T_m$ are recorded at higher temperatures, i.e. at 85 and 88 °C, respectively, with respect to electrospun, non-UV-cured samples F-G5P2 ($T_m$: 75 °C). These results provide additional evidence that the UV-curing treatment successfully lead to the synthesis of a covalent gelatin/PCL co-network within electrospun fibres.

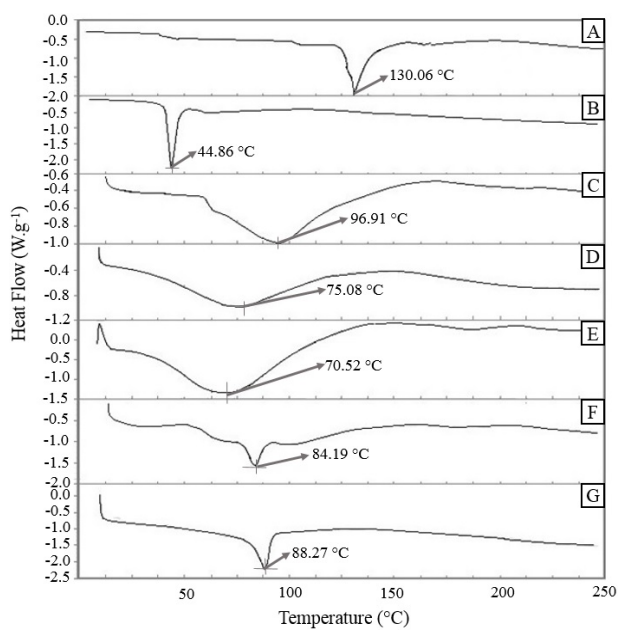

**Fig. 6** Typical DSC thermograms of raw, electrospun and UV-cured samples. (A): G-4VBC; (B): PCL-DMA; (C) F-G30; (D) F-G5P1; (E) F-G5P2; (F) F-G5P2$^L$; (G) F-G5P2$^H$.

*3.5. Elucidation of chemical composition*

ATR-FTIR analysis was carried out to elucidate the chemical composition of electrospun and UV-cured samples (Fig. 7). As expected, the typical absorption bands of gelatin were detected in ATR-FTIR spectra of all samples, i.e. the characteristic peak at: 1632 cm$^{-1}$, related to the stretching vibration of C=O bond (amide I); 1540 cm$^{-1}$ related to the C-N stretching vibrations (amide II) and N-H bending vibrations; 1238 cm$^{-1}$ describing the CH$_2$ wagging vibrations (amide III); and at 3288 cm$^{-1}$ regarding the N–H stretching vibration [92-96]. Other than gelatin-related bands, characteristic absorption bands of the PCL phase could also be identified in electrospun and UV-cured samples at 2949 cm$^{-1}$ and 2865 cm$^{-1}$ (regarding the asymmetric CH$_2$ stretching and symmetric CH$_2$ stretching, respectively), at 1727 cm$^{-1}$ (describing the carbonyl stretching C=O associated with the ester bonds), at 1226 cm$^{-1}$ (related to the C–O and C–C stretching) and at 1185 cm$^{-1}$ (regarding the asymmetric COC stretching) [97-99]. With respect to electrospun samples, ATR-FTIR spectra of UV-cured membranes did not show any remarkable difference, as expected due to the limited detection of vinyl-related bands via FTIR. Overall, the identification of previously-mentioned characteristic absorption bands of PCL in ATR-FTIR spectra confirms the successful incorporation of PCL-DMA in resulting membrane, providing evidence of the key role of PCL crosslinker on the structural fibre stability in hydrated conditions.

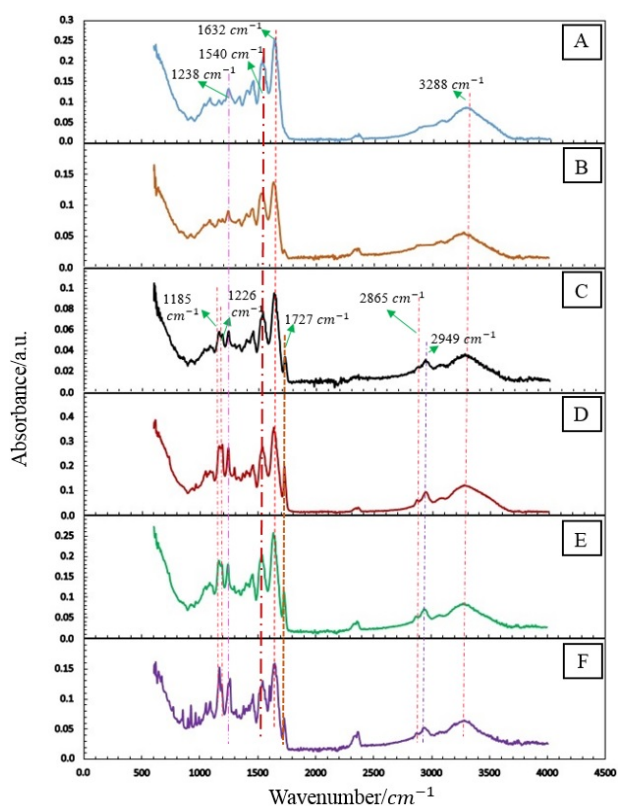

**Fig. 7** ATR-FTIR spectra of samples F-G20 (A), F-G30 (B), F-G5P1 (C), F-G5P2 (D), F-G5P2$^L$ (E), F-G5P2$^H$.

*3.6. Mechanical properties of electrospun and UV-cured membranes*

Figure 8 shows the mechanical properties of electrospun and UV-cured membranes prepared from either gelatin or gelatin/PCL solutions, in both dry and wet conditions, in order to investigate the effect of chemical formulation and molecular architecture on membrane mechanical properties. Corresponding average values of the tensile strength (MPa) and Young's modulus (MPa) were calculated using the linear, plastic and failure regions of the stress-strain curves and are reported in fig. 8 (A-B). Basically, we have considered the average values due to the slightly different values between samples with the same composition, as far as the fact that the mass per unit area of the electrospun nanofibrous membranes in general is not uniform [100]. Generally, the stress strain curves of the electrospun membranes exhibited a linear elastic behaviour at small strains and then a gradual increase in tensile strength at larger strains. In agreement with the fact that PCL electrospun materials have higher tensile strength, lower Young's modulus and higher strain at break than electrospun gelatin [64, 65], introduction of PCL-DMA in resulting fibres led to improved tensile strength and strain at break as well as reduced Young's modulus, as clearly observed by the stress-strain curves of samples F-G5P1 and F-G5P2 (Fig. 8 (C)).

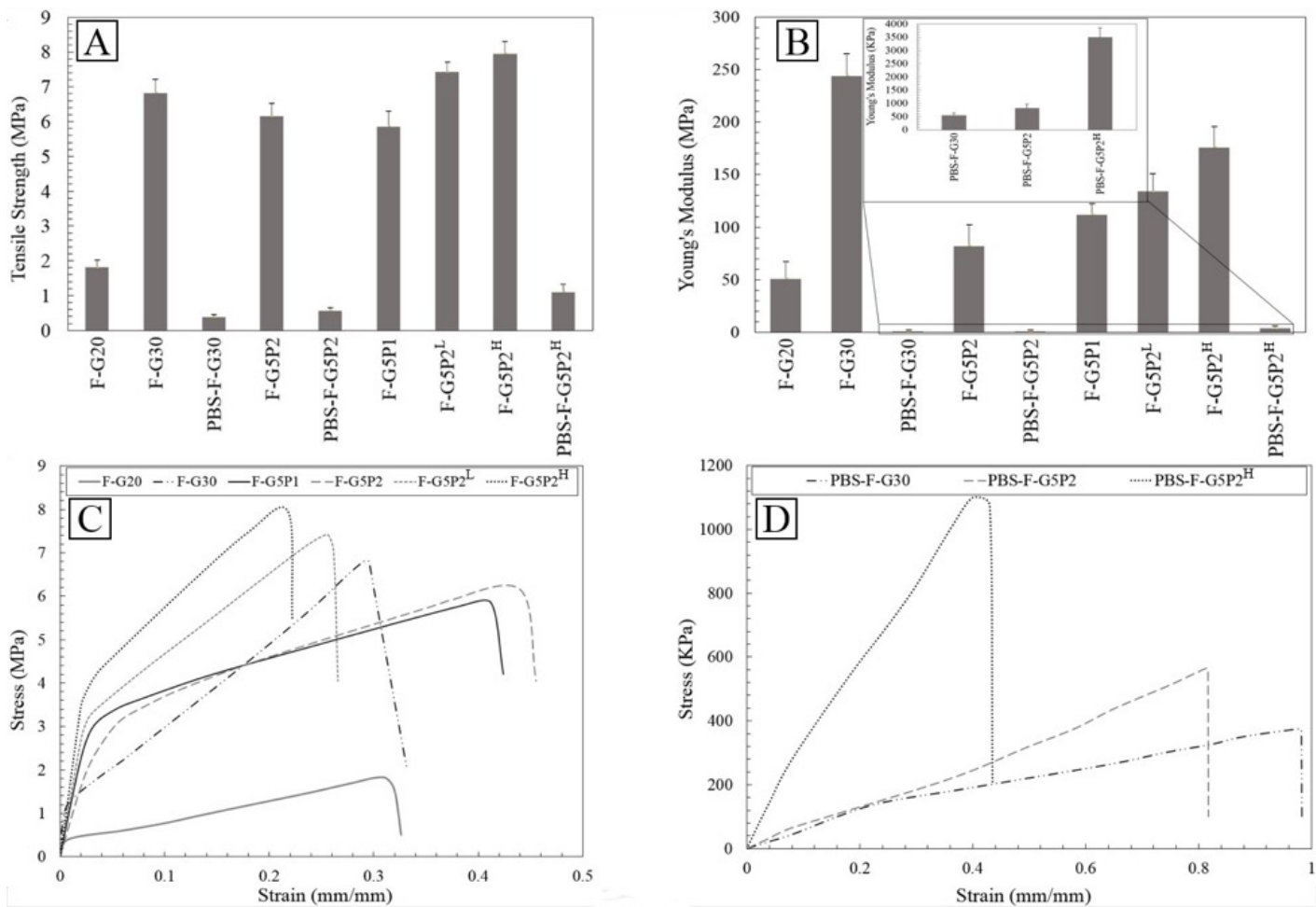

**Fig. 8** Mechanical properties of electrospun and UV-cured membranes. (A) Tensile strength, (B) Young's modulus, (C) Typical tensile stress-strain curves in dry condition, (D) Typical tensile stress-strain curves in wet condition.

It has been proposed that introduction of PCL in a gelatin-based system results in weak physical interactions, less entanglements, higher microphase separation and poor static adhesion amongst the chains of gelatin and PCL [64, 101]. As a result, when tensile loading is applied to respective dry electrospun membrane, increased mobility (i.e. less Young's Modulus) and deformation (i.e. higher strain at break) of the chains are expected. Other that the electrospun state, the tensile stress-strain curves revealed that UV curing improves the mechanical properties of dry gelatin/PCL electrospun samples, again supporting the formation of a covalent gelatin/PCL co-network at the molecular scale. It is clear from the stress strain curves how the UV-induced co-network formation leads to increased tensile strength and Young's modulus and decreased strain at break in samples F-G5P2$^L$ and F-G5P2$^H$ with respect to sample F-G5P2. In fact, the UV-induced crosslinking reaction between fibre-forming gelatin and PCL chains introduces covalent linkages, which lead to reduced chain mobility and fibre strain at break and improved tensile strength and Young's modulus [62, 102]. Moreover, as the concentration of the photoinitiator is increased (e.g. in sample F-G5P2[H] with respect to sample F-G5P2[L]), the consequent generation of additional radicals during UV-curing is expected to induce further yield of photocrosslinking, so that resulting UV-cured membranes will be made of co-networks with increased crosslink density, resulting in increased Young's modulus and tensile strength increased, and decreased strain-at-break. In the wet state (Fig. 8 (D)), the Young's moduli and tensile strengths of electrospun and UV-cured membranes decreased significantly, whilst the strain at break was obviously enhanced compared to those of the membranes in the dry conditions (Fig. 8 (C)). This is mainly attributed due to the presence of large amount of free water dispersed in the pores of the nanofibrous membrane, as well as due to the chemically-bound water present in the fibres, which acted as gelatin plasticiser, thus weakening the interaction among neighbouring fibres [103, 104]. The mechanical properties of the electrospun membranes tested under wet conditions showed similar trends to those tested under dry conditions with regards to the effect of network formation and presence of the hydrophobic PCL-DMA segment. It was reasonably explained that the polymer chains became comparatively uncoiled and separated in the hydrated structure and thus the sliding friction between polymer chains became very low compared to the dry state [105, 106].

*3.7. Wettability characterization of functionalised gelatin and gelatin/PCL composite nanofibrous membranes*

The surface wettability of the nanofibrous membranes was evaluated by time dependent liquid contact angle (CA), since fibre wettability has a great impact on adhesion, spread, proliferation and growth of cells [29, 66, 107]. Methylene Iodide (MI) was used as contact liquid instead of water to avoid water-induced swelling of the gelatin-based membranes and consequent alteration of surface profile [64]. Table 1 reports the images of the contact angles of MI on the surface of nanofibrous membranes produced at four recorded times, revealing an immediate and irregular decrease in contact angle values.

**Table 1** Liquid contact angles and optical images of electrospun and UV-cured membranes recorded at four time points (0, 1, 2 and 3 sec) following application of a methylene iodide (MI) drop.

| Sample | Time = 0 sec | Time = 1 sec | Time = 2 sec | Time = 3 sec |
|---|---|---|---|---|
| F-G5P2[H] | CA (deg) 23° | CA (deg) 13.5° | CA (deg) 11.5° | CA (deg) 10.5° |
| F-G5P2[L] | CA (deg) 21.5° | CA (deg) 6° | CA (deg) 5° | CA (deg) 0° |
| F-G5P2 | CA (deg) 18.5° | CA (deg) 10.5° | CA (deg) 5° | CA (deg) 4.5° |
| F-G5P1 | CA (deg) 16.5° | CA (deg) 9° | CA (deg) 8° | CA (deg) 7.5° |
| F-G30 | CA (deg) 16° | CA (deg) 12° | CA (deg) 8.5° | CA (deg) 3° |

Contact angles measured on UV-cured samples were as high as the ones on electrospun samples, suggesting similar material hydrophilicity despite the formation of the gelatin/PCL co-network. Although PCL nanofibrous membranes are typically hydrophobic with a water contact angle of around $125^0$ [9], the introduction of PCL-DMA in the electrospun gelatin system proved to induce minimal effect on membrane surface wettability, likely related to the low molecular weight of PCL used, and its low content in the electrospinning solution [65].

*3.8. Water holding capacity*

Water holding capacity is an important characteristic of electrospun scaffolds intended for e.g. regenerative medicine and wound healing applications. It can be quantified via gravimetric swelling tests as well by measuring the change in average fibre diameter following contact with water (Fig. 9). The average swelling ratio was decreased from 423 wt.% in sample F-G5P2 to 127 wt.% in respective UV-crosslinked sample F-G5P2$^H$. This trend was clearly confirmed via SEM, whereby the average fibre diameter was more than doubled in water-incubated with respect to dry samples (*d*: 1.3 ±0.2 → 2.9 ± 0.4 μm), whilst nearly no fibre could be retained by either corresponding electrospun sample F-G5P2 or UV-cured samples F-G30$^H$. These results confirm that the UV-induced crosslinking reaction successfully enables controlled swelling of resulting fibres due to the synthesis of the covalent gelatin/PCL co-network [108].

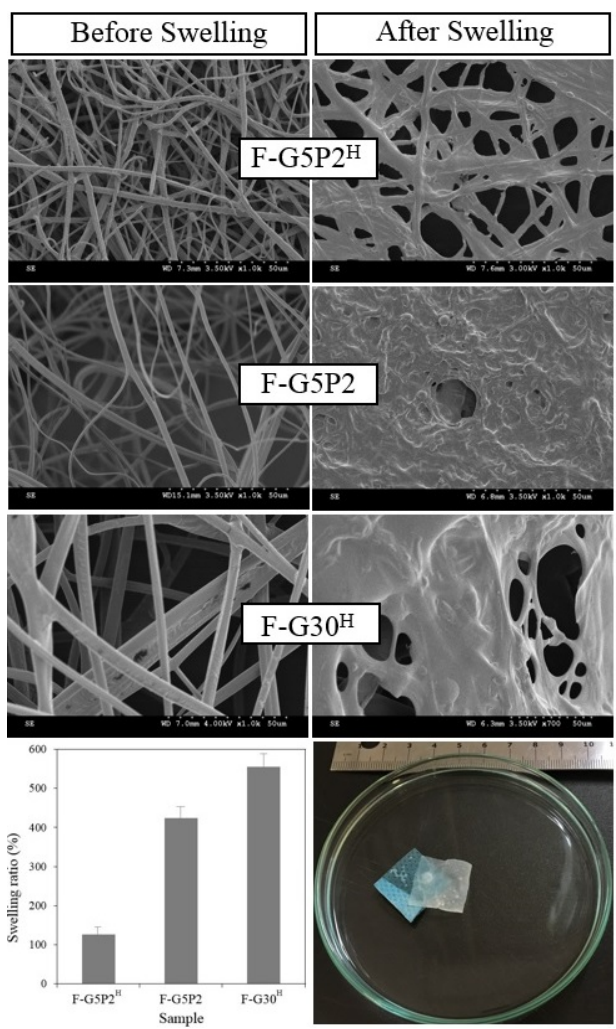

**Fig. 9** SEM images (prior to and following contact with water) and water-induced swelling ratio of electrospun and UV-cured membranes made of either G-4VBC or G-4VBC–PCL-DMA. A photograph of sample F-G5P2$^H$ incubated in water is also presented.

*3.9 In vitro cell culture*

The compatibility of the UV-cured gelatin membrane in biological system was evaluated by investigating the cellular activity of osteosarcoma G292 cells when cultured in direct contact with sample F-G5P2$^H$. Cell metabolism and viability were measured by Alamar blue assay at day 1 and 4 of cell culture (Fig. 10), and confirmed via live /dead staining on cells seeded on sample F-G5P2$^H$ following 4-day culture (Fig. 11). The cellular activity at day 1 recorded on the UV-cured sample was found to be insignificantly different from the cellular activity recorded on the crosslinked gelatin control, crosslinked via state of the art carbodiimide chemistry [77].

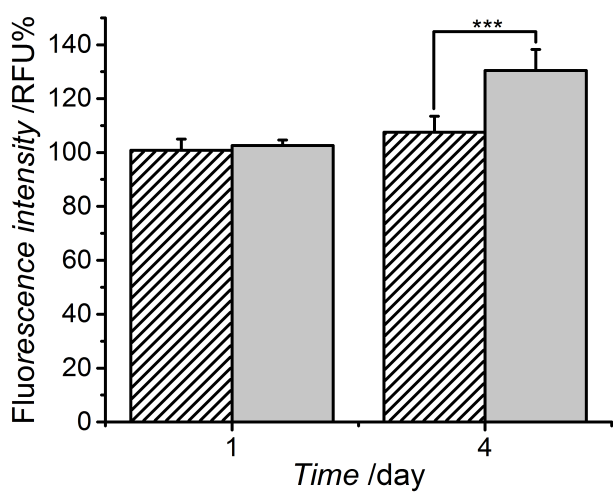

**Fig. 10** Alamar blue assay carried out with G292 osteosarcoma cells over 4 days. Cells were cultured on to either an EDC-crosslinked gelatin control (sparse) or samples F-G5P2$^H$ (light grey). *** indicates significant difference ($p<0.001$, n=6).

On the other hand, cells displayed significantly increased activity at day 4 when cultured on the UV-cured compared to the control sample. In agreement with Alamar blue results, no membrane-induced toxic response was confirmed via live/dead staining of G292 cells following 4-day culture (Fig. 11). These cell culture results therefore successfully demonstrated the high tolerability of the UV-cured gelatin membrane realised in this study with G292 osteosarcoma cells.

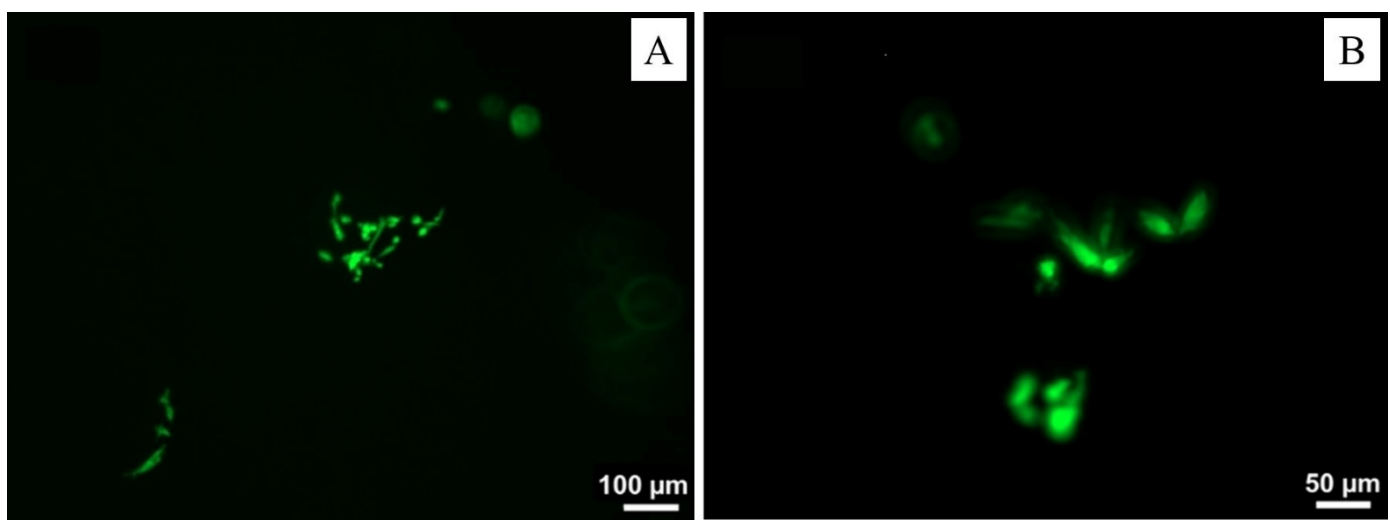

**Fig. 11** Typical live /dead staining of G292 osteosarcoma cells following 4-day culture on sample F-G5P2$^H$ at the magnification of ×10 (A) and ×20 (B), respectively.

## 4. Conclusion

In this study, free surface electrospinning was employed to investigate the scalable manufacture and respective wet-state structural stability of gelatin-based fibrous membranes. 4VBC-functionalised gelatin was used as suitable biomimetic backbone with photopolymerisable groups and dissolved with PCL-DMA, which was applied as hydrophobic, degradable crosslinker. FSES successfully lead to the formation of homogeneous fibres made of a UV-induced covalent co-network at the molecular scale. The employment of Heptanol as photoinitiator carrier for the UV-induced crosslinking reaction played a critical role in order to ensure minimal fibre instability and retained fibrous architecture. Most importantly, the incorporation of PCL-DMA was key towards the synthesis of a covalent co-network enabling controlled water-induced fibre swelling and structural stability in hydrated conditions, whilst also ensuring membrane compatibility with G292 cells *in vitro*. The yield of co-network crosslink density could be adjusted depending on the photoinitiator concentration used to initiate the UV-induced crosslinking reaction, and directly affected the degree of fibre morphology retention in the hydrated state, as well as fibre thermal stability and mechanical properties. Contact angle analysis proved that the hydrophilicity of gelatin/PCL membranes was slightly decreased with respect to gelatin controls, whilst UV-curing did not significantly influence the surface wettability of resulting membranes. This study elucidates the effects of UV-induced crosslinking reaction, co-monomer and photoinitiator towards the chemical stabilisation of gelatin fibres and demonstrates that the presented approach can be applied for the scalable manufacture of fibrous membranes. Next research steps will focus on the material-induced response to skin cells for wound healing applications.

## Acknowledgements

The authors wish to thank the Clothworkers' Centre for Textile Materials Innovation for Healthcare (CCTMIH), the EPSRC-University of Leeds Impact Acceleration Account, and the EPSRC Centre for Innovative Manufacturing in Medical Devices Fresh Ideas Fund for financial support. The University of Leeds, School of Chemical and Process Engineering and School of Physics are also acknowledged for access to contact angle, surface tension and electrical conductivity facilities.

**Supporting Information**

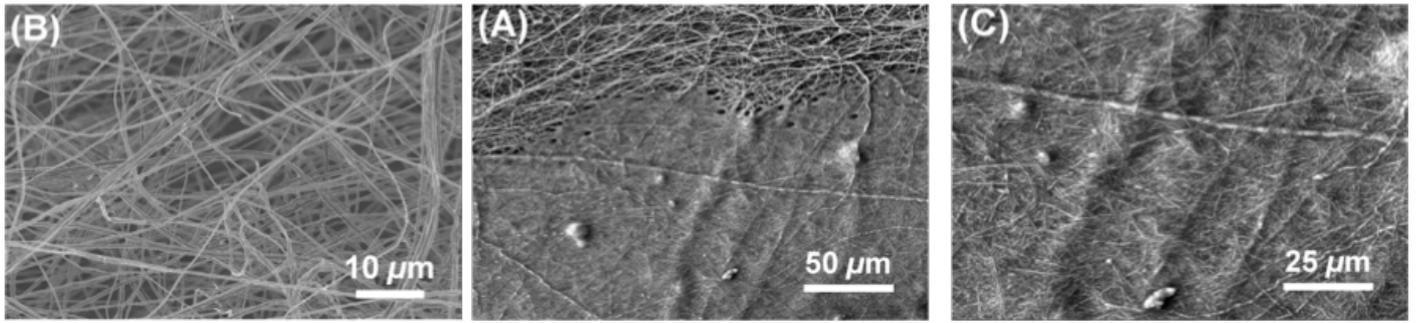

**Figure S1.** SEM images of an electrospun gelatin sample in the dry state (A), and following contact with (B), and complete incubation in (C), ethanol.